\documentclass[preprint,showpacs,preprintnumbers,amsmath,amssymb]{revtex4}

\usepackage{epsf}
\usepackage{epsfig}
\usepackage{graphicx}
\usepackage{dcolumn}
\usepackage{bm}
\begin{document}
\renewcommand{\thesection}{\arabic{section}}
\renewcommand{\thetable}{\Roman{table}}
\setlength{\baselineskip}{16.0pt}
\makeatletter
\renewcommand{\fnum@table}[1]{Table~\thetable. } 
\makeatother

\bibliographystyle{apsrev}

\title{Quantum electron transport in toroidal carbon nanotubes \\ with metallic leads}

\author{
M. Jack$^{\ast}$ and M. Encinosa
\centerline{Florida A\&M University, Department of Physics, 205 Jones Hall, Tallahassee, FL 32307.}
}
\thanks{Corresponding author.  Email: jack@cepast.famu.edu, mencinosa1@comcast.net}


\begin{abstract}
 A recursive Green's function method is employed to calculate the
 density-of-states, transmission function, and current through a
 150 layer (3,3) armchair nanotorus (1800 atoms) with laterally attached metallic
 leads as functions of relative lead angle and magnetic flux.
  Plateaus in the transmissivity through the torus occur over wide
  ranges of lead placement, accompanied by enhancements in the transmissivity through the torus
 as magnetic flux normal to the toroidal plane is varied.

\end{abstract}



\maketitle

\section{Introduction}
Carbon nanotube structures and graphene  materials \cite{iijima,novoselov1,novoselov2,novoselov3,zhang2,peres,pereira}
show promise for advancing  nanoelectronics in the 21st century.
Their unique mechanical, optical and electronic properties make
them important materials for studying  coherent quantum transport
and quantum control of both charge and spin  \cite{shapiro,borzi, potz,qin,oosterkamp,trauzettel,rycerz,hill,son2,son3,son}. Toroidal carbon
nanotube structures, also known as carbon nanotori, could play
interesting roles in nanoelectronics, quantum computing, and as
biosensors \cite{goldsmith,mannik}.  To date, most of the efforts directed towards quantum
rings have focused on flat two-dimensional structures
\cite{liu,bulaev2,climente,datta,datta2,filikhin,fuhrer,georgiev,gylfad,ivanov,latil,pershin1,pershinring,sasaki1,sasaki2,simonin,viefers,maiti,papp}. While these
objects can be more easily modelled in comparison to their
toroidal counter parts, toroidal structures allow for motions
around their minor radius in addition to azimuthal motion, which
could give rise to novel phenomena \cite{encinosa9,encinosa10,encinosa11}. Some of the potential
applications by which the unique transport features of nanoscale
tori might be exploited are:
\begin{itemize}
\item{as macrosopic, three-dimensional molecular rings with
persistent current effects in  magnetic fields or azimuthal and
dipole-type electronic excitations in  microwave fields;} \item{
as molecular Aharonov-Bohm oscillators with modulations in the
current as a function of angle between the attached metallic leads
or magnetic flux parallel to the toroidal symmetry axis;} \item{as
biosensors, in which modifications of device  properties are
altered after covalently attaching a biopolymer.}
\end{itemize}

In this paper, the density-of-states $D(E)$, transmission function
$T(E)$, and  current through the torus $I_{SD}$ between two
attached metallic leads under small  bias are calculated via
tight-binding Green's function methods. The nanotorus modelled
here consists of 150 layers of a (3,3) armchair unit cell giving
1800 carbon atoms. It measures $4\AA$ in  width and has a central
diameter of $D=116\AA$, making it thin compared to the graphene
ring discussed in \cite{recher}  where the ring width-to-radius
$\frac{a}{W}$ is set to $1:5$ or $1:10$ respectively. The opening
angle between the leads is varied along with magnetic flux $\Phi$
through the torus to study oscillations of  $I_{SD}$ due to
constructive or destructive interference of the electronic
pathways through the left and right semi-circular toroidal
branches. Electron-charging effects with band-bending and
modification of the Fermi-level in the torus have been shown to be
small in similar small-bias device architectures utilizing carbon
nanotubes, or can be sufficiently approximated with a linear
potential profile in the device region without major change of
results compared to the exact self-consistent treatment \cite{xue}.
Thus, a full self-consistent treatment of electron-electron
correlation and Coulomb repulsion effects on the quantum transport
 will not be discussed here. The conductivity $G$ and  $I_{SD}$
for the small-bias case can  be obtained from $T(E)$  by a
convolution over $E$ with the difference in the Fermi
distributions functions $f_{L,R}$ of the left and right metallic
leads \cite{anantram}.
%

This paper is organized as follows: In section 2 the Green's
function formalism in the tight-binding approximation by which the
results obtained here are presented.
 In section 3 graphs of $T(E)$ and $D(E)$ and of $I_{SD}$ as functions of different small-bias source-drain
voltages $V_{SD}$ are shown.  A magnetic field parallel to the
central symmetry axis ($z$-axis)
 is applied to study  $I_{SD}$ as a function of magnetic flux through the structure.
 Section 4 is reserved for conclusions and suggestions for further work.

\section{ Transport in the tightbinding approximation}
The  device setup for the toroidal nanotube is sketched in Figure
1. Each semi-infinite metallic leads touches the toroidal surface
at four atomic contact points.  The positions of these contacts can be
determined  from the toroidal vector $\mathbf{r}$
\begin{equation}
 \mathbf{r} (\theta,\varphi)= (R + a \ {\rm cos} \theta ){\mathbf
e}_{\rho} +a\  {\rm sin} \theta {\mathbf e}_{z}
\end{equation}
and the relative positions of the two leads. Effectively, the
curvature of the torus is neglected at the lead interface. The
opening angle $\alpha$ between the two leads can be varied from
$180^{\mbox{o}}$ ({\it back-to-back configuration}) down to
$45^{\mbox{o}}$ (or $315^{\mbox{o}}$ respectively).

The Hamiltonian for a nanotube in the
 tight-binding approximation is \cite{anantram}:
\begin{equation}
 H = \sum\limits_i{E_i c_i^{\dagger} c_i}  + \sum\limits_{i>j}\left( t_{ij} c_i^{\dagger} c_j  + {h.c.} \right)
\end{equation}
with $E_i$ the  on-site energies and the $t_{ij}$ the hopping
terms. Here all  $E_i$=0 and the $t_{ij}$ for the torus will be
set equal to $v = -3.1\,\mbox{eV}$ in accordance with the
literature \cite{dresselhaus}. These assignments neglect any curvature effects
on the $t_{ij}$.

 The  device is modelled as an open system under a small bias voltage using a
 Green's function method \cite{caroli,meir,charlier,anantram}.
 The semi-infinite leads can be absorbed into the definition of a device Green's function $G_d$
  that serves to yield transport properties of the nanotorus.
  Self-energy corrections $\Sigma_{L,R}$ describing the coupling of the device
  region to both leads can be folded into a device Hamiltonian $H_d$ that carries the retarded
  and advanced GreenÕs functions $G_d^{(a,d)}(E)$ as solutions to a unit
  source via

\begin{equation}
 \left[  E - H_d  -\Sigma_L - \Sigma_R \pm  i \eta \right]  G_d^{(a,r)} = I.
\end{equation}
The Hamiltonian  $H_d$ is an $1800\times 1800$ matrix at the C-C
level but may be re-written as a $150\times 150$ matrix whose
elements consist of $12 \times 12$ intra or interlayer armchair
coupling submatrices $A$ and $V$ respectively,
\begin{eqnarray}
\label{matrix}
 H_d =
 \left(
 \begin{array}{ccccccc}
A & -V^{\dagger} & 0 & \ldots & \ldots & 0 & -V
\\
-V & A & -V^{\dagger} & 0 & \ldots & 0 & 0
\\
0 & -V & \ddots & \ddots  & & & 0
\\
\vdots & 0 &\ddots &&&& \vdots
\\
\vdots & \vdots &&&& \ddots & 0
\\
0 & \vdots &&&\ddots &\ddots & -V^{\dagger}
\\
-V^{\dagger} & 0 & \ldots & \ldots & 0 & -V & A
 \end{array}
 \right)
 \end{eqnarray}

A numerically fast, recursive algorithm to invert the bracketed
term in Eq.(3) to compute $G_d$ is known for an infinite nanotube.
For the toroidal carbon nanotorus, however, there is an additional
complication due to the corner elements in the Hamiltonian needed to
achieve ring closure. An algorithm that extends the simple case to
include toroidal ring closure has been developed by one of us  \cite{encinosa12}.
and is employed here to determine the device Green's function.
 $\Sigma_{L,R}$ can be calculated  analytically for semi-infinite metallic
  leads.  All relevant transport
  properties can be derived from the retarded and advanced Green's functions
  $G_d^{(a,d)}(E)$ of the device region.

 The GreenÕs function of the semi-infinite leads, $g_m$, can be
determined analytically. Discretizing along the longitudinal
direction with lattice spacing $a$ as is standard practice \cite{datta2,anantram}
gives for a lead with cross-sectional area $L_yL_z$
\begin{equation}
g_m(E) = {{-8 m a} \over {\hbar}^2}{1 \over{ L_y L_z}}  \sum
      sin {{m \pi y} \over L_y} sin {{n \pi z} \over L_z}
      sin {{m \pi y^\prime} \over L_y} sin {{n \pi z^\prime} \over L_z}
    [\sqrt{{{\alpha^2}_{mn}}-1}  - \alpha_{mn}],
\end{equation}
\begin{equation}
\alpha_{mn} = {E+E_F  - E_{mn} \over t} -1,\qquad  t = {\hbar^2
\over m a^2},
\end{equation}
with $E_{mn}$ the standard particle in a box eigenvalue expression
and $E_F$ the Fermi energy of the metal, included to account for
the zero of energy being set at the tube Fermi energy $E=0$  \cite{anantram}.
Eq.(5)  is understood to be evaluated on the layer closest to
the torus.

The exact value of the corresponding electronic coupling
$t_{\mbox{hop}}$ of the first atomic layer of the metallic lead
surface to the torus surface is strongly dependent on the choice
of the metal and its lattice structure. $t_{\mbox{hop}} =
-0.25\,\mbox{eV}$ is chosen as initial default \cite{anantram} and its value
later modified and its effect on $T(E)$ later studied.

The formula for $D(E)$, $T(E)$ and $I_{SD}$ under bias are
summarized below.  For this, the coupling terms $\Gamma_{L,R}(E)$
for the lead-torus connections are relevant and computed from the
self energies $\Sigma_{L,R}$ and subsequently from $g_m(E)$:
\begin{equation}
\Gamma_k(E) = 2\pi i \left[ \Sigma_k(E)- \Sigma^{\dagger}_k(E) \right] =
2\pi V^{\dagger}_k Im\left[  g^k_m(E) \right] V_k; k = L, R.
\end{equation}

From Eq.(3), (4) and (7), $D(E)$ and transmission $T(E)$
are determined as:
%
%
\begin{equation}
D(E) = G_d \left( \Gamma_L + \Gamma_R \right) G^{\dagger}_d;
\end{equation}
%
%
\begin{equation}
T(E) = \textrm{Trace}\left[   \Gamma_L  G_d \Gamma_R G^{\dagger}_d \right].
\end{equation}
The integrated source-drain current $I_{SD}$ is finally given as:
\begin{equation}
I_{SD} = \frac{2e}{h} \int\limits_{-\infty}^{+\infty} d{E} \, T(E)
\left[ f_D\left( E - \mu_L\right) - f_D\left( E - \mu_R\right)
\right]
\end{equation}
with $f_D(E)={1 \over [\exp(E-\mu)+1]}$.

Finally, we are also interested in studying coherence in electron
transport induced by  electrons interacting with an external
magnetic field. The vector potential
$\mathbf{A}\left(\mathbf{r}\right)$ for a static magnetic field
aligned with the toroidal symmetry axis can be written in the
Coulomb gauge as:
\begin{equation}
\mathbf{A}\left(\mathbf{r}\right) = \frac{1}{2} B_0 \rho {\mathbf
e}_{\varphi}\rightarrow \mathbf{B} = B_0 {\mathbf e}_z.
\end{equation}
 It can be shown that $\mathbf{A}\left(\mathbf{r}\right)$
 can be incorporated  into the $H_d$  in terms of a
 phase factor $\zeta$ via
\begin{equation}
v \rightarrow  v \exp\left[  \frac{i e}{\hbar} \mathbf{A}
\left(\mathbf{r}\right)\cdot\left( \mathbf{r}_i - \mathbf{r}_j
\right)  \right].
\end{equation}

\section{Results}
%
\subsection{Density-of-states $D(E)$ and transmission function $T(E)$}
In Figure 2, density-of-states $D(E)$ and transmission function
$T(E)$ plots are depicted for different choices of $B_0$ and
$t_{\mbox{hop}}$.  In Figures 2a and 2b, results for $D(E)$ and $T(E)$ are compared
at $B_0=0$,  $B_0=0.5\,\mbox{T}$ and $B_0=1.5\,\mbox{T}$. In each
case the overall  peak structures of $D(E)$ and $T(E)$ stay
essentially unchanged. Van Hove peaks \cite{dresselhaus} are clearly enhanced
with respect to the $B_0=0$ case. For different $B_0$, different
peaks are enhanced. $D(E)$ has values typically in the range $\sim
10^{-2}\,\mbox{eV}^{-1}$ to $\sim 10^{2}\,\mbox{eV}^{-1}$ with
enhancements up to $\sim 10^{3}\,\mbox{eV}^{-1}$ for energies $E =
-1.0\,\mbox{eV}$ to $E = 1.0\,\mbox{eV}$.  $T(E)$ ranges between
$\sim 10^{-10}$ and $\sim 10^{-2}$ for the same energy interval.
The symmetry of $D(E)$ and $T(E)$ for positive and negative
energies is apparent.
The hopping parameters at the metal-torus contacts and in the
device region were set to  $t_{\mbox{hop}} = -0.25\,\mbox{eV}$ and
$v = -3.1\,\mbox{eV}$.  The Fermi level in metallic leads is chosen
at $E_{\rm fermi} =  6.0 \,\mbox{eV}$ in rough accordance with
physical values.

Part (c) clearly demonstrates that improving the electronic coupling to the metallic contacts causes an increase in number of available states at the contacts into and out of which electrons can be scattered from and/or into the device region, as evidenced  by a positive shift of the density-of-states $D(E)$ with an increase in $t_{\mbox{hop}}$.  $t_{\mbox{hop}}$ is modified here from the default value $t_{\mbox{hop}} = -0.25 \,\mbox{eV}$ to $t_{\mbox{hop}} = -0.5 \,\mbox{eV}$ and $t_{\mbox{hop}} = -1.5 \,\mbox{eV}$ respectively.  Again, values for $D(E)$ are in the range $\sim 10^{-2}\ldots 10^{3}\,\mbox{eV}^{-1}$. Similarly, an overall symmetric increase in $T(E)$ is observed in Figure 2d when increasing $t_{\mbox{hop}}$ allowing close to unit $T(E)\approx 1$, for certain energies.
%

\subsection{Coherence in electronic transmission}
Up to now, results for $D(E)$ and $T(E)$ have been presented with
the metallic leads attached in a {\it back-to-back} configuration,
i.e. the opening angle $\alpha$ is chosen as $180^{\mbox{o}}$.
 Having in mind  analogous phenomena seen in flat annular structures,
 the question as to whether the torus could act
 as a macro-molecular Aharonov-Bohm oscillator arises.
 When changing the opening angle $\alpha$ between the leads and applying
 a bias voltage across leads, the effective path length of electrons moving
 in the left or right branch of the nanotorus from metallic lead to lead
 will differ, opening up possible interference effects between the left and
 right electron wave functions.  Due to the quasi-ballistic nature of  transport
  in carbon nanotube-like structures, coherence in transmission through
  the nanotorus should be expected.
   The three-dimensional geometry of the nanotorus and its
    finite (non-zero) width will show a modulation in the oscillations
    of $T(E)$ for different $\alpha$ observed in rings \cite{liu,bulaev2,climente,datta,filikhin, fuhrer,georgiev,gylfad,ivanov,latil,pershin1,pershinring,sasaki1,sasaki2,simonin,viefers,maiti,papp} due to
    differing possible electronic pathways with
    different major radii varying between $R-a$ and $R+a$ in the toroidal
    geometry.

The recursive algorithm employed here easily allows for the
modelling of a rotation of the position of one of the leads with
respect to the other by determining the new positions of the four
atomic contact positions for the rotated lead while the positions
of the contact sites at the other fixed lead remain unchanged.   A
quick  back-of-the-envelope estimate helps to determine the
electron momenta $k$ and thus for which de Broglie wavelengths
$\lambda=\frac{2 \pi}{k}$ coherence in electronic transmission is
visible before the finite size of the lattice constant of the
underlying graphene lattice of $a=1.4 \AA$ washes out any
interference effects for higher momenta.
 The electronic energy $E$ is roughly given by $E=\frac{\hbar^2 k^2}{2 m_e}$.  Assuming $E\approx 0.01\mbox{eV}$, calculate the occurence of interference minima in transmission
 in steps of $\Delta\alpha$ where destructive interference occurs for
  pathlength differences $\Delta s = \left( 2 n + 1 \right) \frac{\lambda}{2}$
   with $n = 0, 1, 2, \ldots$ between the left and right toroidal branch:
\begin{eqnarray}
\left.
 \begin{array}{ccc}
E &\approx & 0.01\,\mbox{eV} = \frac{\hbar^2 k^2}{2 m_e} \rightarrow \lambda \approx 12.3\,\mbox{nm}
\\
R &=& 5.8\mbox{nm} \rightarrow u = 2\pi R = 36.4\,\mbox{nm}
 \end{array}
\right\}
\rightarrow
\Delta\alpha \approx 360^{\mbox{o}} \cdot \frac{\lambda / 2}{u} = 60.6^{\mbox{o}}.
\end{eqnarray}

The minimum angular difference at which the finite lattice size
starts playing a role for a (3,3) armchair carbon nanotorus with
1800 atomic sites and for the dimensions used here is $\Delta\alpha =
\frac{360^{\mbox{o}}}{150} = 2.4^{\mbox{o}}$.  Thus, for
electronic energies $E$ of $\sim  0.1{\mbox{eV}}$ and larger, any
coherence effects visible at lower energies will become
negligible.

The above predicted plateaus in  $T(E)$ due to constructive
interference in  transmission through the torus can be clearly
deduced from Figures 3a and 3b.  $T(E)$ was calculated for $E=0.01\,\mbox{eV}$ and $E=0.02\,\mbox{eV}$ where plateaus according
  to Eq.(13) were predicted.  The figures show several interference curves for an angular
  range of $\Delta\alpha = 45^{\mbox{o}}, \ldots, 315^{\mbox{o}}$.  At larger energies,
  calculations show a gradual flattening of the plateaus and eventually a disappearance
   of coherence effects at $E=0.1\,\mbox{eV}$ and $E=0.2\,\mbox{eV}$ respectively.
    At smaller energies on the order of $0.001 \mbox{eV}$ or less, $T(E)$
     appears essentially flat with small amplitudes  $\sim  10^{-4}$ and less,
     quickly decreasing when turning on and increasing $B_0$.
$B$-field values from $0\,\mbox{T}$ to larger fields of
$5.0\,\mbox{T}$ have been chosen to determine any dependence of
the position of the plateaus as a function of   flux threading the
nanotorus plane. A substantial enhancement of $T(E)$ by about one
order of magnitude occurs for specific $B_0$ values
 $B_0\approx 1.0 \ldots 1.5\,\mbox{T}$;  the exact
$B$-field for maximum $T(E)$ plateau shifts to larger values with
increasing energy $E$.

 For comparison, the $T(E)$ plots for non-resonant magnetic field values are shown enhanced in
 the insets.  The resonant $B$-field dependence of the transmission
plateaus at the higher scale
 corresponds to the internal atomic scale of the  torus, $a=1.4 \AA$.
 In this case, the magnetic phase $\zeta=e^{i \Phi / \Phi_0}$ with $\Phi_0 = {h\over e}$ creates
   a $B$-field dependent electronic interference on top of the previously discussed interference effect
   due to different opening angles $\alpha$.  In simple terms, constructive
   and destructive interferences are generated by a magnetic phase change $\zeta$
    that accrue from lattice site to lattice site along the toroidal circumference to yield
    overall macroscopic phase changes for coherence in transmission from the left
    and right toroidal branches while at the larger scale this macroscopic
    phase change already occurs for electron hopping from lattice site to adjacent
    lattice site.  Further details can be found in Subsection 3.3.

\subsection{$I_{SD}$ with magnetic flux; the torus as an Aharonov-Bohm oscillator}
\label{aharonov}

Under a small voltage bias $V_{SD}$, $I_{SD}$ can be determined from Eq.(10) by an
integration over the transmission function $T(E)$
 convolved with the difference in the Fermi distributions of the
 two metallic leads. (Here the effect of the field on the leads has been
 neglected but should be considered in more comprehensive work.)  $I_{SD}$ is plotted as a
 function of  $V_{SD}$ in Figures 4a and 4c for a number
 of different $B_0$ while Figures 4b and 4d show the dependency
  of $I_{SD}$ on $\Phi$ for two $V_{SD}$ values.

Linear voltage-current profiles for the small-bias case can be
clearly shown in Figures 4a and 4c for back-to-back leads and
$90^{\mbox{o}}$ lead opening angle.  The slope of the curves
yields the conductance $G \times V_{SD}$ with $G$ as conductivity
as predicted from Landauer-B{\"u}ttiker theory in small-bias
approximation \cite{landauer,buttiker}. The
 current can change by a factor of roughly 2.5 when choosing $B_0 = 0, 0.5, 1.0, 1.5, 2.0 \mbox{T}$.
  From this depiction no pattern in the
dependency of $I_{SD}$ on $B_0$ can be detected.

Following the discussion of coherence patterns in transmission for different opening angles between
 leads, one could conceive  of achieving similar interference effects by increasing  magnetic
 flux $\Phi = \int\limits_c {\mathbf A} \, d{\mathbf r}$  through
 the torus.
The flux ratio $\frac{\Phi}{\Phi_0}$ with respect to  $\Phi_0$ can
be equated with a magnetic field ratio $\frac{B}{B_0}$ with
$B_0=0.026 \,\mbox{T}$.  The current $I_{SD}$ is
plotted here as a function of the flux ratio
$\frac{\Phi}{\Phi_0}$. For both the $180^{\mbox{o}}$ and
$90^{\mbox{o}}$ lead configurations, similar oscillations of
$I_{SD}$ with changing magnetic flux are demonstrated. In each
case, the strongly oscillatory curve appears to be composed of a
superposition of two simpler oscillations in $\Phi$ not shown here
with $\Phi$ as even or odd multiples of
 $\frac{1}{16}\Phi_0$.  Constructive and destructive interference
 in electronic transmission can thus be generated due to phase angle
 difference with magnetic fluxes at the scale of $\frac{\Phi_0}{16}$.
  The flux period of these oscillations can be read off from the plots
  as $\Delta\Phi=\frac{3}{16}\Phi_0$.  For comparison,
   magnetic flux dependencies are mainly computed and shown
   in the literature for persistent currents for two- and three-dimensional ring
    geometries \cite{pershin1,pershinring,recher,sasaki1}, but the magnitude of $I_{SD}$ for example is
    comparable to the current magnitude expected in a two-dimensional
    graphene ring under similar conditions when re-scaling the results in \cite{recher}
    to the torus dimensions here.  Also, the general oscillatory behavior with changing
     magnetic flux appears to agree with findings reported there and other places \cite{bulaev2,climente,datta,filikhin, fuhrer,georgiev,gylfad,ivanov,latil,pershin1,pershinring,sasaki1,sasaki2,simonin,viefers,maiti,papp}.

To compute $I_{SD}\left( V_{SD} \right)$, $T(E)$ is calculated symmetrically around the device Fermi level
$E^d_f$ between the energies $E = -0.2\mbox{eV}$ and $E = 0.2\mbox{eV}$ for a step size of
  $5\times 10^{-5}\mbox{eV}$ where $E^d_f$ has been set to zero.  $T(E)$ is then
  integrated over this energy region with the difference in Fermi distributions
  according to Eq.(10) to approximately yield $I_{SD}$.  For the small-bias
  case $V_{SD} = 0, \ldots, 0.1\mbox{eV}$, contributions to the energy integration
  of $I_{SD}$ beyond the chosen energy range may be safely neglected due to
  the  fast exponential drop-off of the Fermi functions.

\section{Conclusions}
The authors employ a recursive Green's function algorithm to
calculate the electronic density-of-states $D(E)$ and transmission
functions $T(E)$ of a thin toroidal carbon nanotube under a small
bias voltage
 between laterally attached metallic leads.

A strong enhancement of $D(E)$ and $T(E)$ is observed when the
nanotorus is threaded with a magnetic flux $\Phi$, particularly
near van Hove peaks \cite{dresselhaus}.  The enhancement is pronounced with
increasing magnetic field.  An increase of the hopping parameter
$t_{\mbox{hop}}$, which governs the effective electronic coupling
at the metal-nanotorus atomic contact sites, also shifts
$D(E)$ and $T(E)$ for all energies.

Coherent electronic interference phenomena could be observed in
 $T(E)$ as a function of the angle $\alpha$ between the metallic leads, and in the
 integrated source-drain current $I_{SD}$ as a function of magnetic flux. $T(E)$ shows
 clear maximum and minimum plateaus for specific angular ranges where constructive or destructive
 interference of the electronic wave functions for electrons moving left or right in the torus
 from one lead to the other can be expected depending on the opening angle $\alpha$.
Due to finite size effects in the nanotorus (i.e. the torus has a non-zero width)
the constructive interference regions are broadened to plateaus over a range of angles
and not simply peak-like maxima.  Similarly, constructive and destructive
 electron interference appears in $I_{SD}$ when applying $B_0$
  where effective path length differences are induced by a magnetic phase angle $e^{i\Phi / \Phi_0}$,
  different for the left and right toroidal branches.
$I_{SD}$ measured in units of $\frac{e}{h}$ shows strong oscillations in $\Phi / \Phi_0$.
 The magnitude of $I_{SD}$ is comparable to the current expected in a two-dimensional
 graphene ring under similar conditions after re-scaling the results in \cite{recher} to the toroidal dimensions.

Natural  extensions of the model and methods of this work are towards larger systems with
more than tens of thousands of carbon atoms, and to include nearest
and next-to-nearest neighbor contributions.  For flat two-dimensional graphene nanoribbons,
an opening of an energy bandgap has been predicted \cite{son}.  Similar effects can be expected to appear in an expanded calculation for the three-dimensional geometries.
Furthermore, a more realistic device setup with the metallic leads
attached at the $\lq$bottom' of the torus, i.e. the torus lieing flat
on the two planar leads will allow an increase in the number of
metal-carbon contact sites and possibly lead to better and closer
lead attachment. This could prove to be very interesting  in the
context of an applied microwave  field with an additionally tilted
static magnetic field due to the interplay between electronic
excitations and  persistent and bias-driven currents through the
torus \cite{encinosa9,encinosa10,encinosa11}.  It is conceivable that quantum control applications could
follow when the induced dipole and solenoidal magnetic moments
couple to electronic spin.

\begin{center} {\bf Acknowledgments} \end{center}
One of the authors (M.E.) would like to thank M.P. Anantram and M. Meyyapan
for initially suggesting this project and  for helpful support by the NASA Ames Research Center.
 M.J. is currently being supported under NSF Grant 0603070.  Both authors would also like to
 thank L. Johnson and N. Christopher at Florida A\&M University's Laser Remote Sensing Laboratory
  (LRSL) Computer Cluster.



\newpage

\begin{center} {\bf Figure captions} \end{center}

\vskip 24pt \noindent Figure 1.
Sketch of device setup with toroidal carbon nanotube and metallic leads.

\vskip 24pt \noindent Figure 2.
{\it Density-of-states $D(E)$ and transmission function $T(E)$ as a function of energy $E$:}
 a. $D(E)$ for different magnetic fields $B_0$.
 b. $T(E)$ for different magnetic fields $B_0$.
 c. $D(E)$ for different electronic coupling terms $t_{\mbox{hop}}$ between torus and metallic leads.
 d. $T(E)$ for different hopping terms $t_{\mbox{hop}}$.

\vskip 24pt \noindent Figure 3.
{\it Coherence in electronic transport.  Plateaus in electronic transmission:}
a. Transmission function $T(E)$ at $E=0.01\,\mbox{eV}$ as a function of magnetic field $B$
for different angles $\alpha$ between metallic leads.
b. Transmission function $T(E)$ at $E=0.02\,\mbox{eV}$.

\vskip 24pt \noindent Figure 4.
{\it Back-to-back leads:}  a. Source-drain current $I_{SD}$ as a function of source-drain
 voltage $V_{SD}[\,\mbox{eV}]$ (small bias) for different magnetic fields $B_0$.
  $I_{SD}$ in units of $\frac{e}{h}$.
  Chemical potential at left/right lead: $\mu_{1,2}= \pm \frac{V_{SD}}{2}$.
  Thermal energy: $k_B T = 30\,\mbox{meV}$.
  b. Source-drain current $I_{SD}$ as a function of applied magnetic field
  $B_0[\,\mbox{T}]$ ($V_{SD}=0.1\,\mbox{eV}$).  {\it $90^o$ angle btw. leads:}
   c. Source-drain current $I_{SD}$versus source-drain voltage $V_{SD}[\,\mbox{eV}]$.
   d. Source-drain current $I_{SD}$ versus magnetic field $B_0[\,\mbox{T}]$.

\newpage

\begin{center} {\bf Figures} \end{center}

\begin{figure}[htb]
\label{devicesketch}
\begin{center}
\vspace{0cm}
\hspace{0cm}
\mbox{\epsfxsize=8.0cm
  \epsffile{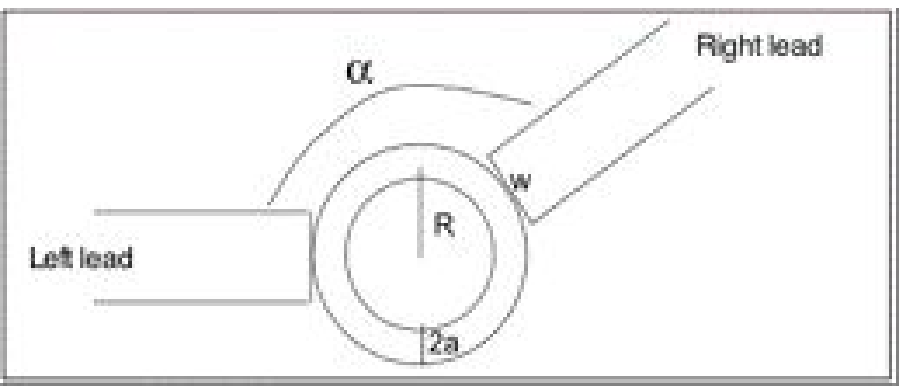}
}
\end{center}
\centerline{Fig. 1.}
\end{figure}
%
%
\begin{figure}[htb]
\label{de-te-plots}
\begin{minipage}[htb]{8cm}
{
\begin{center}
  \vspace{0cm}
  \hspace{-1cm}
  \mbox{
    \epsfxsize=8.5cm
        \epsffile{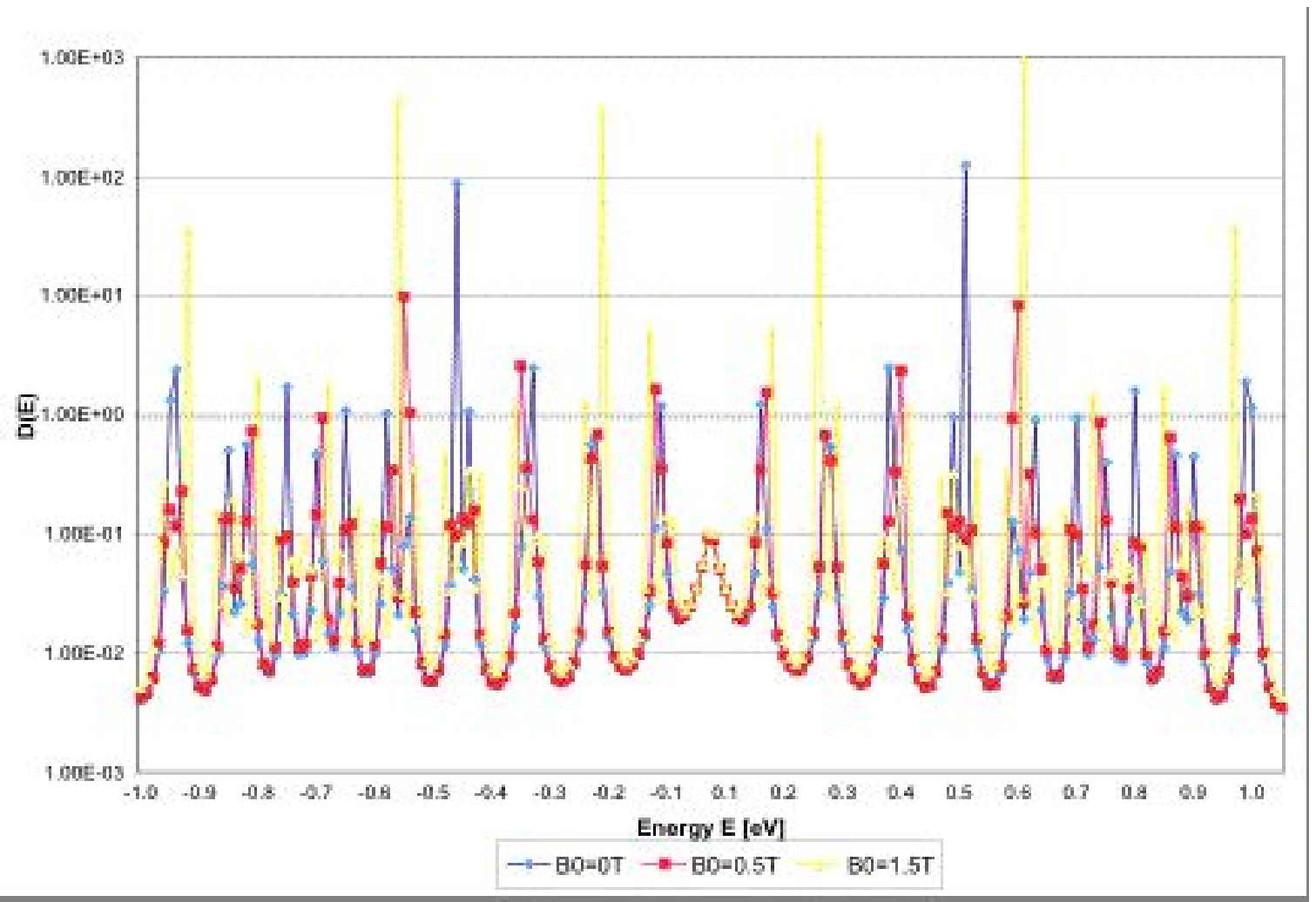}
  }
 \end{center}
 \centerline{Fig. 2 a.}
 }
\end{minipage}
\begin{minipage}[htb]{8cm}
{
\begin{center}
  \vspace{0cm}
  \hspace{0cm}
  \mbox{
    \epsfxsize=8.5cm
        \epsffile{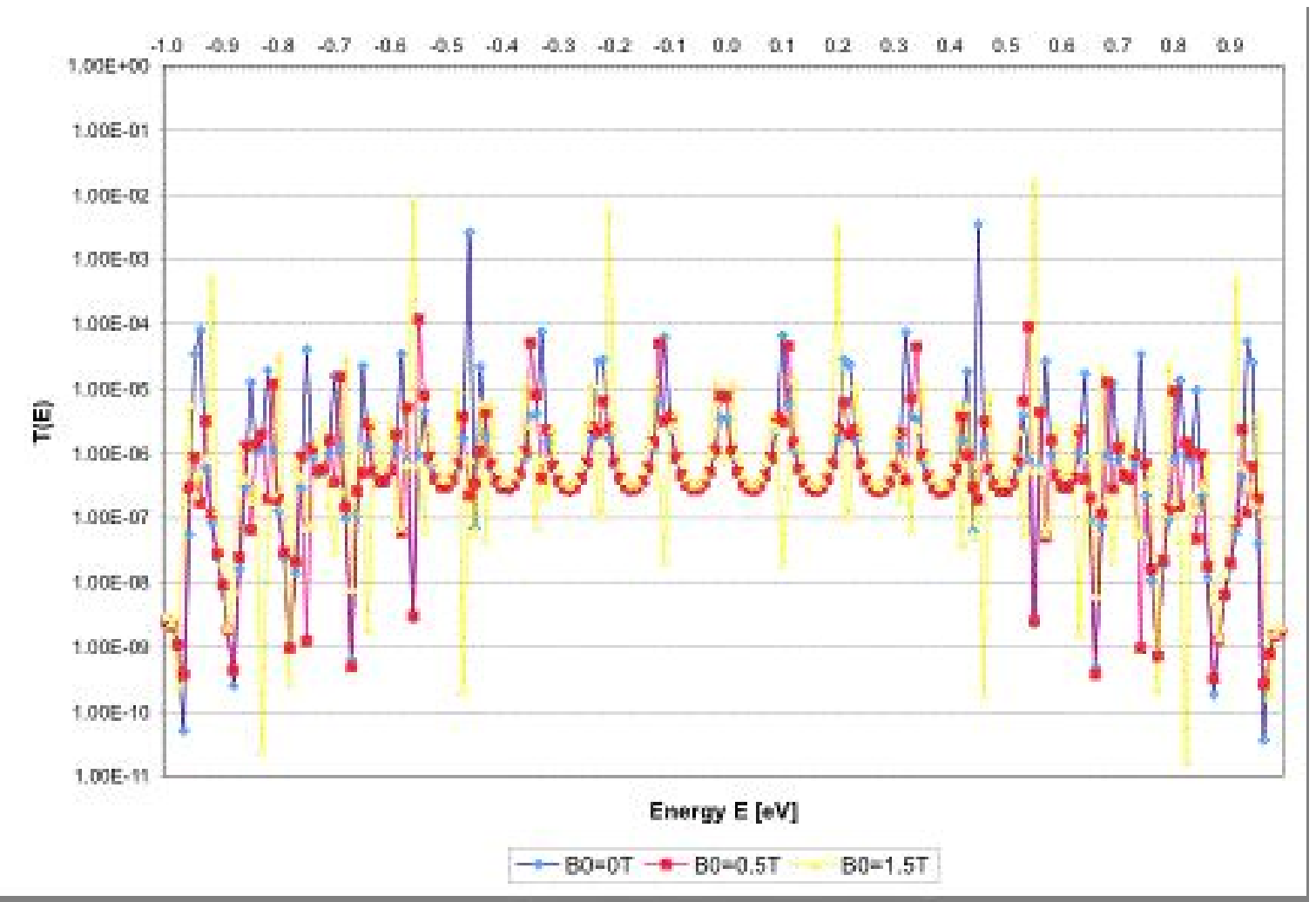}
  }
 \end{center}
 \centerline{Fig. 2 b.}
 }
\end{minipage}
\begin{minipage}[hbt]{8cm}
{
\begin{center}
  \vspace{0.5cm}
  \hspace{-1cm}
  \mbox{
    \epsfxsize=8.5cm
        \epsffile{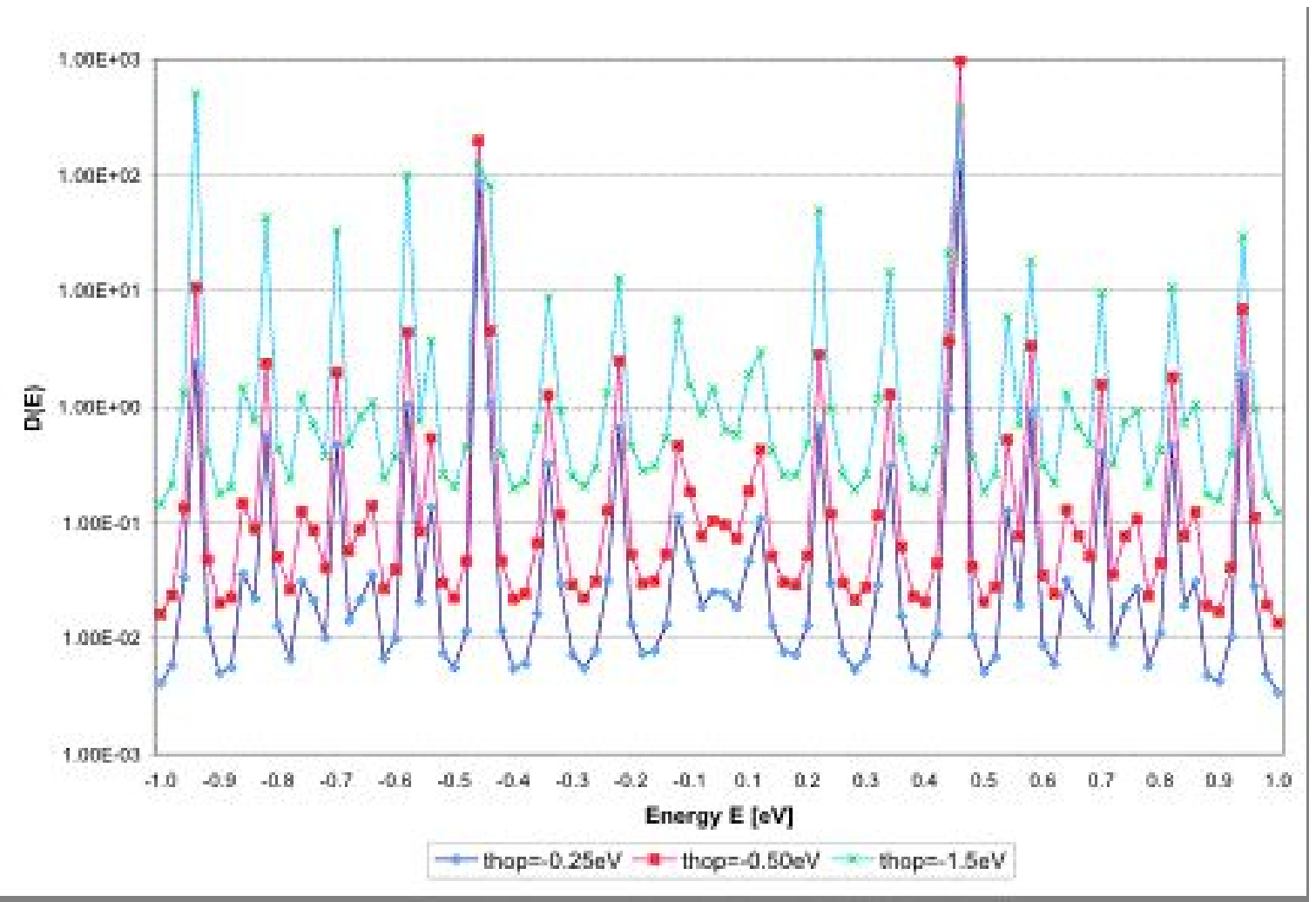}
  }
 \end{center}
 \centerline{Fig. 2 c.}
 }
\end{minipage}
\begin{minipage}[hbt]{8cm}
{
\begin{center}
  \vspace{0.5cm}
  \hspace{0cm}
  \mbox{\epsfxsize=8.5cm
        \epsffile{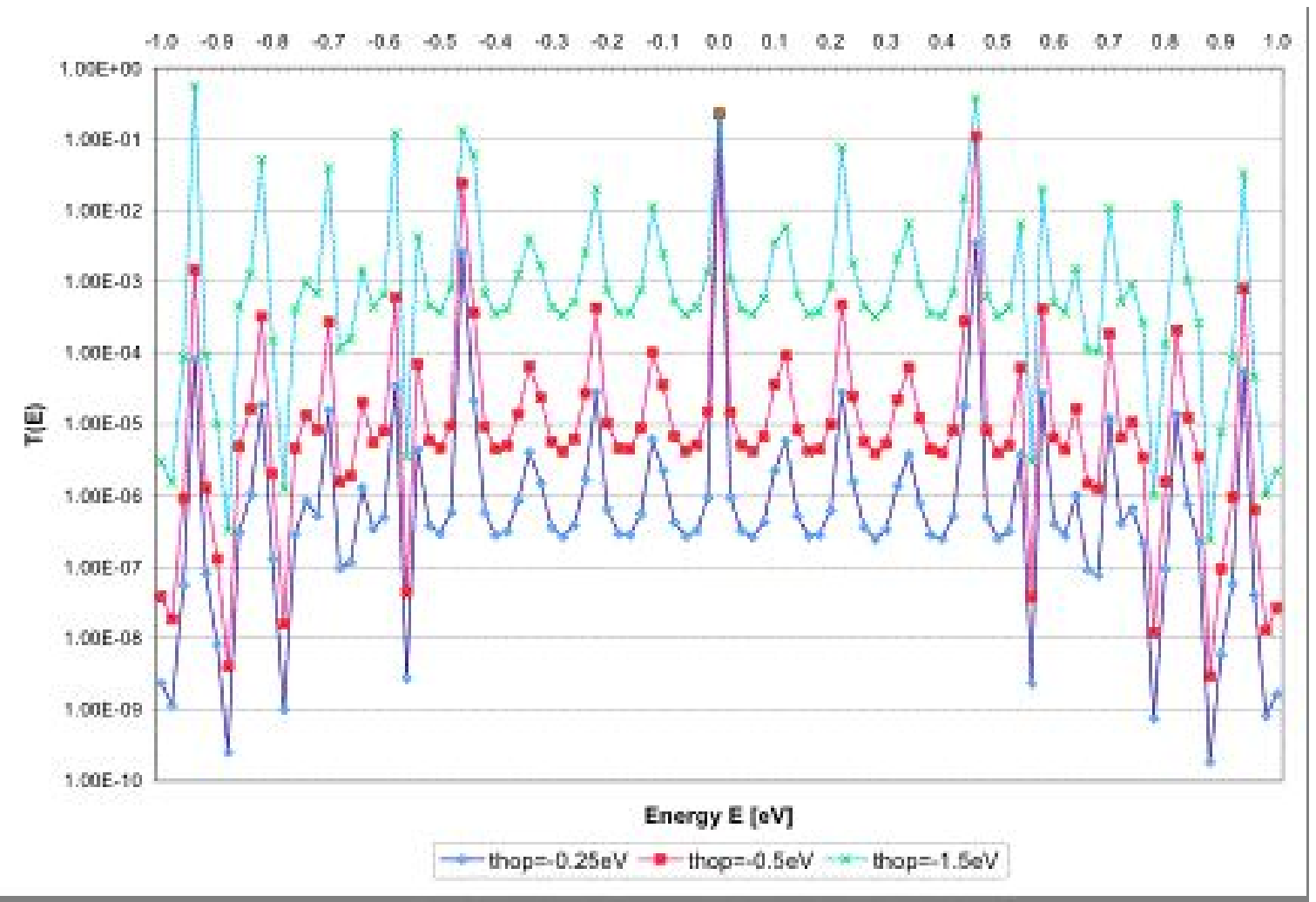}
    }
 \end{center}
 \centerline{Fig. 2 d.}
 }
\end{minipage}
\end{figure}
%

\begin{figure}[htb]
\label{plateauplots}
\begin{minipage}[bht]{8cm}
{
\begin{center}
  \vspace{0cm}
  \hspace{-1.0cm}
  \mbox{\epsfxsize=8.5cm
        \epsffile{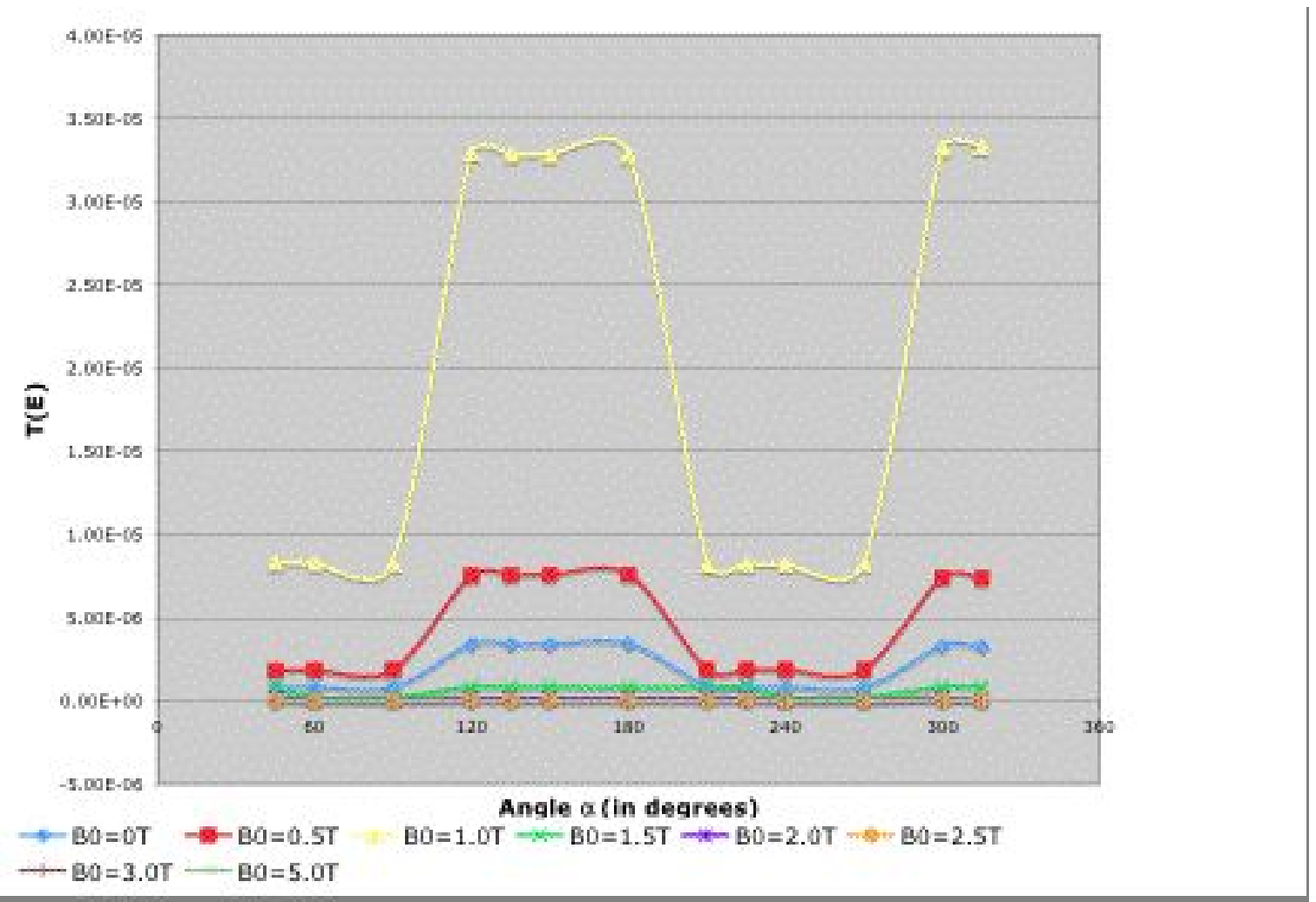}
  }
 \end{center}
 \centerline{Fig. 3 a.}}
\end{minipage}
\begin{minipage}[bht]{8cm}
{
\begin{center}
  \vspace{0cm}
  \hspace{1cm}
  \mbox{\epsfxsize=8.5cm
        \epsffile{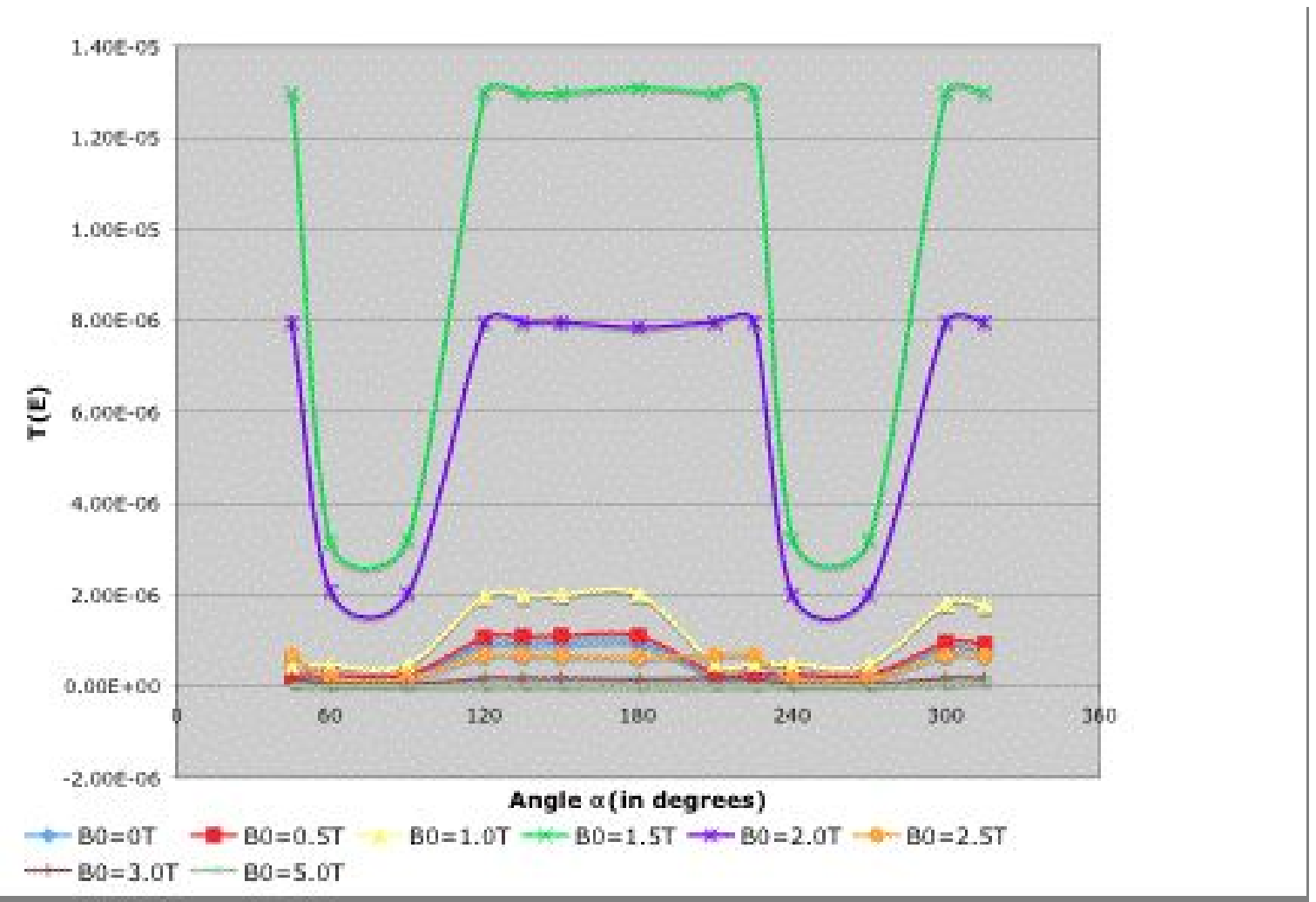}
  }
 \end{center}
 \centerline{Fig. 3 b.}
 }
\end{minipage}
\end{figure}
%

\begin{figure}[htb]
\label{currentplots}
\begin{minipage}[htb]{8cm}
{
\begin{center}
  \vspace{0cm}
  \hspace{-1.0cm}
  \mbox{
    \epsfxsize=8.5cm
        \epsffile{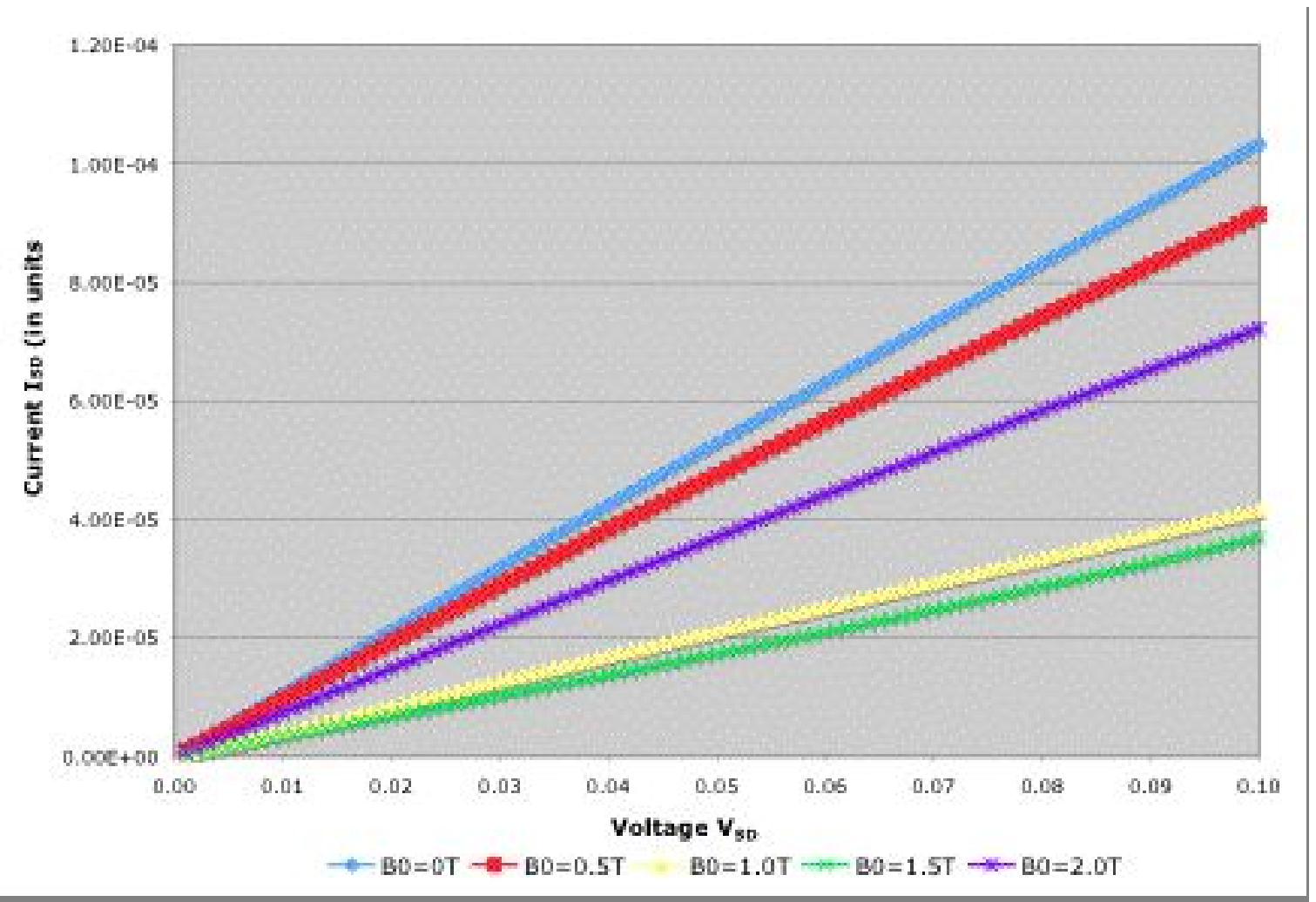}
  }
 \end{center}
 \centerline{Fig. 4 a.}
 }
\end{minipage}
\begin{minipage}[htb]{8cm}
{
\begin{center}
  \vspace{0cm}
  \hspace{0cm}
  \mbox{
    \epsfxsize=8.5cm
    \epsffile{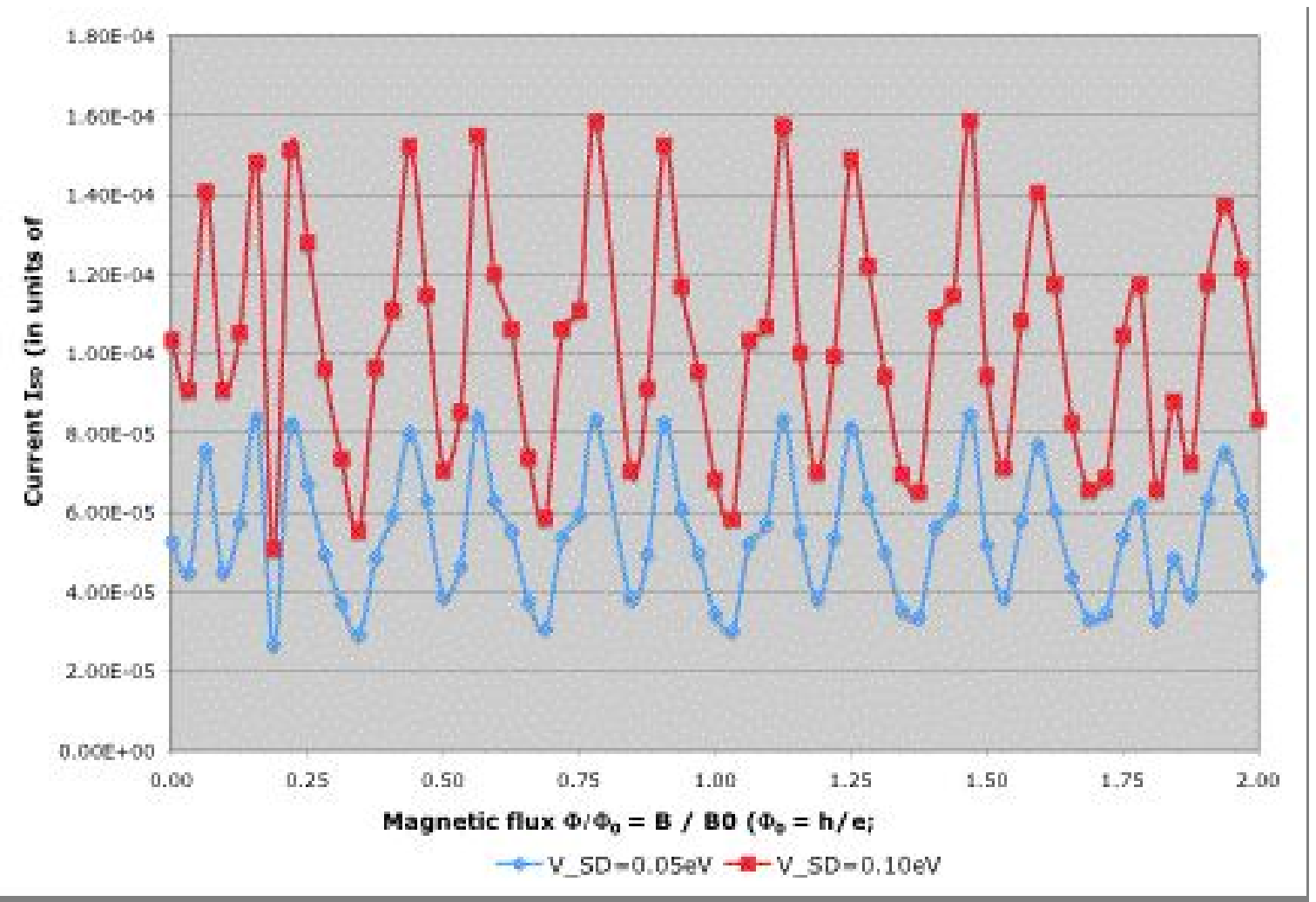}
  }
 \end{center}
 \centerline{Fig. 4 b.}
 }
\end{minipage}
\begin{minipage}[htb]{8cm}
{
\begin{center}
 \vspace{0.5cm}
  \hspace{-1.0cm}
  \mbox{
    \epsfxsize=8.5cm
        \epsffile{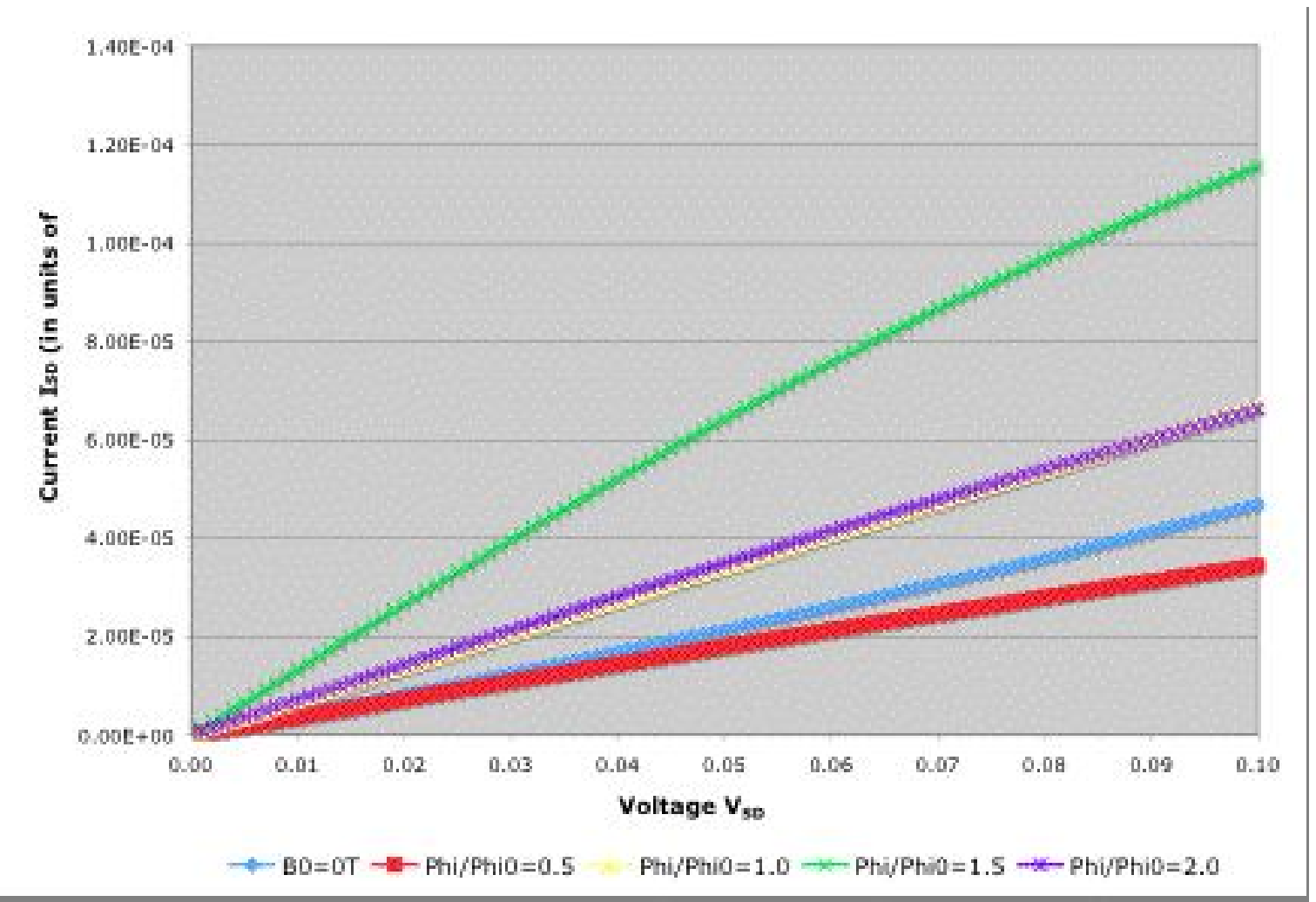}
    }
 \end{center}
 \centerline{Fig. 4 c.}
 }
\end{minipage}
\begin{minipage}[htb]{8cm}
{
\begin{center}
 \vspace{0.5cm}
  \hspace{0cm}
  \mbox{
    \epsfxsize=8.5cm
        \epsffile{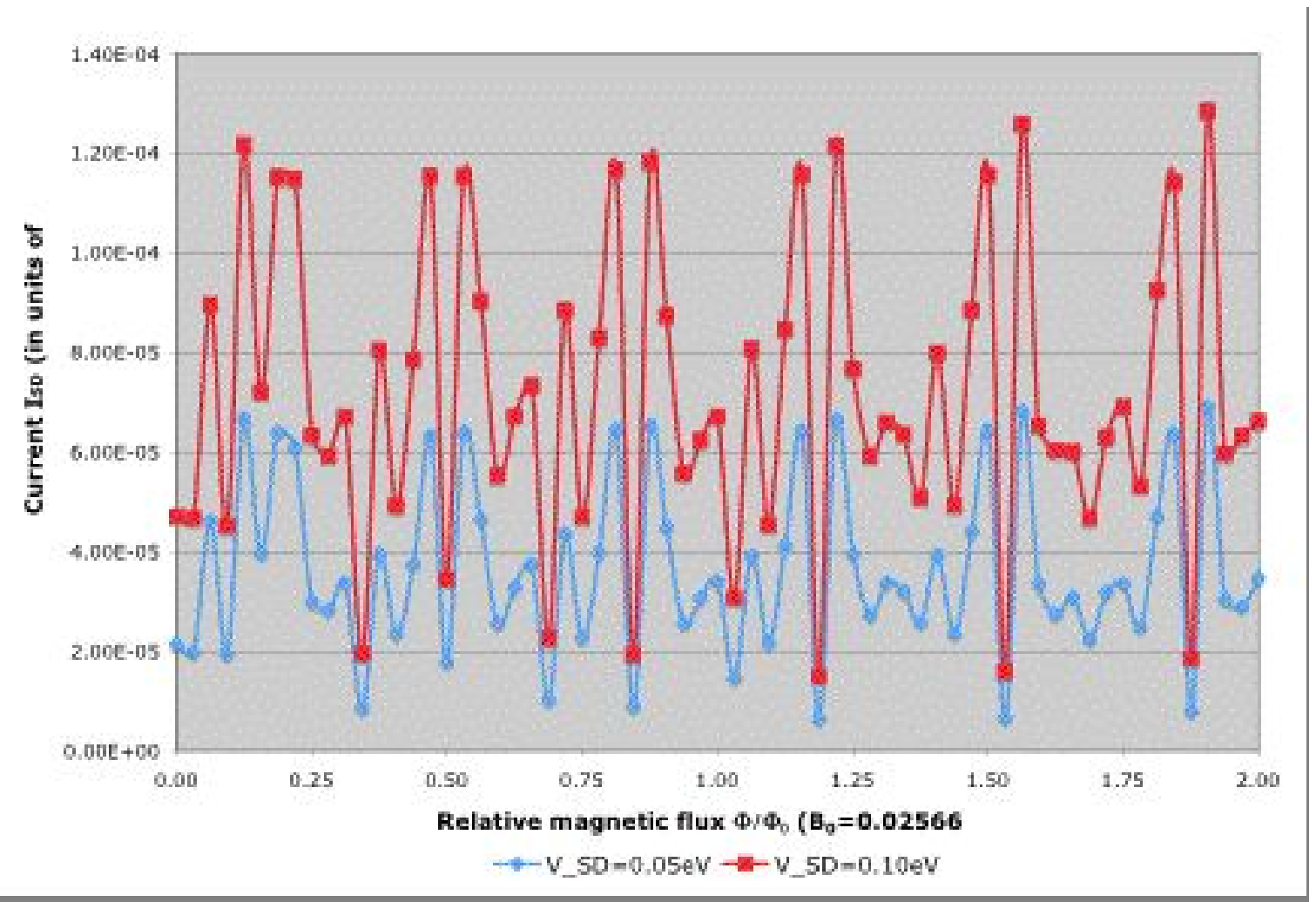}
    }
 \end{center} 
 \centerline{Fig. 4 d.}
 }
\end{minipage}
\end{figure}

\end{document}